\documentclass[prl,showpacs,twocolumn,superscriptaddress]{revtex4}
\usepackage{amssymb}
\usepackage{epsfig}
\usepackage{graphicx,amsmath}
\begin{document}

\title{Behavior of second order phase transitions at a quantum critical point}
\author{V.R. Shaginyan}\email{vrshag@thd.pnpi.spb.ru}
\affiliation{Petersburg Nuclear Physics Institute, RAS, Gatchina,
188300, Russia}\affiliation{Racah Institute of Physics, Hebrew
University, Jerusalem 91904, Israel}
\author{M.Ya. Amusia}\affiliation{Racah Institute
of Physics, Hebrew University, Jerusalem 91904, Israel}
\author{K.G. Popov}\affiliation{Komi Science Center, Ural Division,
RAS, Syktyvkar, 167982, Russia}

\begin{abstract}
Low-temperature specific-heat measurements on $\rm YbRh_2Si_2$ at
the second order antiferromagnetic (AF) phase transition reveal a
sharp peak at $T_N=$ 72 mK. The corresponding critical exponent
$\alpha$ turns out to be $\alpha=0.38$, which differs significantly
from that obtained within the framework of the fluctuation theory of
second order phase transitions based on the scale invariance, where
$\alpha\simeq0.1$. We show that under the application of magnetic
field the curve of the second order AF phase transitions passes into
a curve of the first order ones at the tricritical point leading to
a violation of the critical universality of the fluctuation theory.
This change of the phase transition is generated by the fermion
condensation quantum phase transition. Near the tricritical point
the Landau theory of second order phase transitions is applicable
and gives $\alpha\simeq1/2$. We demonstrate that this value of
$\alpha$ is in good agreement with the specific-heat measurements.
\end{abstract}

 \maketitle

Fundamental understanding of the low-temperature physical properties
of such strongly correlated Fermi systems as heavy fermion (HF)
metals in the vicinity of a quantum phase transition persists as one
of the most challenging objectives of condensed-matter physics. List
of these extraordinary properties are markedly large. Recent
exciting measurements on $\rm YbRh_2Si_2$ at the second order
antiferromagnetic (AF) phase transition extended the list and
revealed a sharp peak in low-temperature specific heat, which is
characterized by the critical exponent $\alpha=0.38$ and therefore
differs drastically from those of the conventional fluctuation
theory of second order phase transitions \cite{TNsteg}, where
$\alpha\simeq0.1$ \cite{land1}. The obtained large value of $\alpha$
casts doubts on the applicability of the conventional theory and
sends a real challenge for theories describing the second order
phase transitions in HF metals.

The striking feature of the fermion condensation quantum phase
transition (FCQPT) is that it has profound influence on
thermodynamically driven second order phase transitions provided
that these take place in the non-Fermi liquid (NFL) region formed by
FCQPT \cite{ams,obz}. For example, the second order superconducting
phase transition in $\rm CeCoIn_5$ changes to the first one in the
NFL region \cite{shag1}. As we shall see, it is this feature that
gives the key to resolve the challenge.

It is a common wisdom that low-temperature and quantum fluctuations
at quantum phase transitions form the specific heat, magnetization,
magnetoresistance etc., which are drastically different from that of
conventional metals \cite{senth,col,lohneysen,si,sach}. Usual
arguments that quasiparticles in strongly correlated Fermi liquids
"get heavy and die" at a quantum critical point commonly employ the
well-known formula basing on assumptions that the $z$-factor (the
quasiparticle weight in the single-particle state) vanishes at the
points of second-order phase transitions \cite{col1}. However, it
has been shown that this scenario is problematic \cite{khodz}. On
the other hand, facts collected on HF metals demonstrate that the
effective mass strongly depends on temperature $T$, doping (or the
number density) $x$, applied magnetic fields $B$ etc, while the
effective mass $M^*$ itself can reach very high values or even
diverge, see e.g. \cite{lohneysen,si}. Such a behavior is so unusual
that the traditional Landau quasiparticles paradigm does not apply
to it. The paradigm says that elementary excitations determine the
physics at low temperatures. These behave as Fermi quasiparticles
and have a certain effective mass $M^*$ which is independent of $T$,
$x$, and $B$ and is a parameter of the theory \cite{land}. A concept
of FCQPT preserving quasiparticles and intimately related to the
unlimited growth of $M^*$ had been developed in Refs.
\cite{khs,ams,volovik}. In contrast to the Landau paradigm based on
the assumption that $M^*$ is a constant, the FCQPT approach supports
an extended paradigm, the main point of which is that the
well-defined quasiparticles determine the thermodynamic and
transport properties of strongly correlated Fermi-systems, $M^*$
becomes a function of $T$, $x$, $B$, while the dependence of the
effective mass on $T$, $x$, $B$ gives rise to the non-Fermi liquid
behavior \cite{obz,khodb,zph,ckz,plaq}. Studies show that the
extended paradigm is capable to deliver an adequate theoretical
explanation of the NFL behavior in different HF metals and HF
systems \cite{obz,khodb,ckz,plaq,shag1,shag2,shag3}.

In this letter, we analyze the specific-heat measurements on $\rm
YbRh_2Si_2$ in the vicinity of the second order AF phase transition
with $T_N=72$ mK \cite{TNsteg}. The measurements reveal that the
corresponding critical exponent $\alpha=$ 0.38 which differs
drastically from that produced by the fluctuation theory of second
order phase transitions, where $\alpha\simeq$ 0.1. We show that
under the application of magnetic field $B$ the curve $T_N(B)$ of
the second order AF phase transitions in  $\rm YbRh_2Si_2$ passes
into a curve of the first order ones at the tricritical point with
temperature $T_{cr}=T_N(B_{cr})$. This change is generated by FCQPT.
Near the tricritical point the Landau theory of second order phase
transitions is applicable and gives $\alpha\simeq1/2$ \cite{land1}.
This value of $\alpha$ is in good agreement with the specific-heat
measurements. As a result, we conclude that the critical
universality of the fluctuation theory is violated at the AF phase
transition due to the tricritical point.

We start with visualizing the main properties of FCQPT. To this end,
consider the density functional theory for superconductors (SCDFT)
\cite{gross}. SCDFT states that the thermodynamic potential $\Phi$
is a universal functional of the number density $n({\bf r})$ and the
anomalous density (or the order parameter) $\kappa({\bf r},{\bf
r}_1)$ and provides a variational principle to determine the
densities \cite{gross}. At the superconducting transition
temperature $T_c$ a superconducting state undergoes the second order
phase transition. Our goal now is to construct a quantum phase
transition which evolves from the superconducting one. In that case,
the superconducting state takes place at $T=0$ while at finite
temperatures there is a normal state. This means that in this state
the anomalous density is finite while the superconducting gap
vanishes. For the sake of simplicity, we consider a homogeneous
Fermi (electron) system. Then, the thermodynamic potential reduces
to the ground state energy $E$ which turns out to be a functional of
the occupation number $n({\bf p})$ since $\kappa=\sqrt{n(1-n)}$
\cite{dft,gross,yakov,plaq}. Upon minimizing $E$ with respect to
$n({\bf p})$, we obtain
\begin{equation}\label{FCM}
\frac{\delta E}{\delta n({\bf p})}=\varepsilon({\bf
p})=\mu,\end{equation} where $\mu$ is the chemical potential. It is
seen from Eq. \eqref{FCM} that instead of the Fermi step, we have
$0\leq n(p)\leq 1$ in certain range of momenta $p_i\leq p\leq p_f$
with $\kappa$ is finite in this range. Thus, the step-like Fermi
filling inevitably undergoes restructuring and formes the fermion
condensate (FC) as soon as Eq. \eqref{FCM} possesses not-trivial
solutions at some point $x=x_c$ when $p_i=p_f=p_F$
\cite{khs,obz,khodb}. Here $p_F$ is the Fermi momentum and $x
=p_F^3/3\pi^2$.

At any small but finite temperature the anomalous density $\kappa$
(or the order parameter) decays and this state undergoes the first
order phase transition and converts into a normal state
characterized by the thermodynamic potential $\Phi_0$. At $T\to0$,
the entropy $S=-\partial \Phi_0/\partial T$ of the normal state is
given by the well-known relation \cite{land}
\begin{equation}
S_0=-2\int[n({\bf p}) \ln (n({\bf p}))+(1-n({\bf p})\ln (1-n({\bf
p}))]\frac{d{\bf p}}{(2\pi) ^3},\label{SN}
\end{equation}
which follows from combinatorial reasoning. Since the entropy of the
superconducting ground state is zero, it follows from Eq. \eqref{SN}
that the entropy is discontinuous at the phase transition point,
with its discontinuity $\Delta S=S_0$. The heat $q$ of transition
from the asymmetrical to the symmetrical phase is $q=T_cS_0=0$ since
$T_c=0$. Because of the stability condition at the point of the
first order phase transition, we have $\Phi_0[n({\bf
p})]=\Phi[\kappa({\bf p})]$. Obviously the condition is satisfied
since $q=0$.

At $T=0$, a quantum phase transition is driven by a nonthermal
control parameter, e.g. the number density $x$. To clarify the role
of $x$, consider the effective mass $M^*$ which is related to the
bare electron mass $M$ by the well-known Landau equation \cite{land}
\begin{equation}\label{LANDM}
\frac{1}{M^*}=\frac{1}{M}+\int \frac{{\bf p}_F{\bf p_1}}{p_F^3}
F({\bf p_F},{\bf p}_1)\frac{\partial n(p_1,T)}{\partial p_1}
\frac{d{\bf p}_1}{(2\pi)^3}.
\end{equation}
Here we omit the spin indices for simplicity, $n({\bf p},T)$ is
quasiparticle occupation number, and $F$ is the Landau amplitude. At
$T=0$, Eq. \eqref{LANDM} reads \cite{pfit,pfit1}
\begin{equation}\label{MM*}
\frac{M^*}{M}=\frac{1}{1-N_0F^1(x)/3}.\end{equation} Here $N_0$ is
the density of states of a free electron gas and $F^1(x)$ is the
$p$-wave component of Landau interaction amplitude $F$. When at some
critical point $x=x_c$, $F^1(x)$ achieves certain threshold value,
the denominator in Eq. \eqref{MM*} tends to zero so that the
effective mass diverges at $T=0$ \cite{pfit,pfit1}. It follows from
Eq. \eqref{MM*} that beyond the critical quantum point $x_c$, the
effective mass becomes negative. To avoid unstable and physically
meaningless state with a negative effective mass, the system must
undergo a quantum phase transition at the quantum critical point
$x=x_c$, which is FCQPT \cite{khs,ams,obz,khodb}.

\begin{figure} [! ht]
\begin{center}
\vspace*{-0.5cm}
\includegraphics [width=0.49\textwidth]{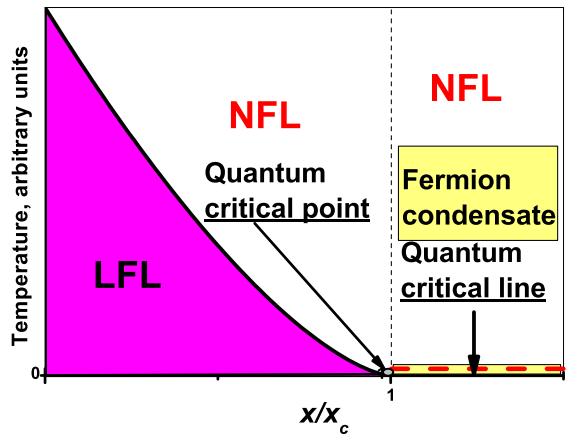}
\end{center}
\vspace*{-0.8cm} \caption{Schematic phase diagram of the system
driven to the FC state. The number density $x$ is taken as the
control parameter and depicted as $x/x_c$. The quantum critical
point, $x/x_c=1$, of FCQPT is shown by the arrow. At $x/x_c<1$ and
sufficiently low temperatures, the system in the Landau Fermi liquid
(LFL) state as shown by the shadow area. At $T=0$ and beyond the
critical point, $x/x_c>1$, the system is at the quantum critical
line depicted by the dash line and shown by the vertical arrow. The
critical line is characterized by the FC state with finite
superconducting order parameter $\kappa$. At $T_c=0$, $\kappa$ is
destroyed, the system undergoes the first order phase transition and
exhibits the NFL behavior at $T>0$.}\label{fig1}
\end{figure}

Schematic phase diagram of the system which is driven to FC by
variation of $x$ is reported in Fig. \ref{fig1}.  Upon approaching
the critical density $x_c$ the system remains in LFL region at
sufficiently low temperatures \cite{khodb,obz}, that is shown by the
shadow area. At the quantum critical point $x_c$ shown by the arrow
in Fig. \ref{fig1}, the system demonstrates the NFL behavior down to
the lowest temperatures. Beyond the critical point at finite
temperatures the behavior is remaining the NFL one and is determined
by the temperature-independent entropy $S_0$ \cite{yakov}. In that
case at $T\to 0$, the system is approaching a quantum critical line
(shown by the vertical arrow and the dashed line in Fig. \ref{fig1})
rather than a quantum critical point. Upon reaching the quantum
critical line from the above at $T\to0$ the system undergoes the
first order quantum phase transition, which is FCQPT taking place at
$T_c=0$.

At $T>0$ the NFL state above the critical line, see Fig. \ref{fig1},
is strongly degenerated, therefore it is captured by the other
states such as superconducting (for example, by the superconducting
state in $\rm CeCoIn_5$ \cite{shag1,shag2,yakov}) or by AF state
(e.g. AF one in $\rm YbRh_2Si_2$ \cite{plaq}) lifting the
degeneracy. The application of magnetic field $B>B_{c0}$ restores
the LFL behavior, where $B_{c0}$ is a critical magnetic field, such
that at $B>B_{c0}$ the system is driven towards its Landau Fermi
liquid (LFL) regime \cite{shag2}. In some cases, for example in HF
metal $\rm CeRu_2Si_2$, $B_{c0}=0$, see e.g. \cite{takah}, while in
$\rm YbRh_2Si_2$, $B_{c0}\simeq 0.06$ T \cite{geg}. In our simple
model $B_{c0}$ is taken as a parameter.

\begin{figure}[!ht]
\begin{center}
\includegraphics [width=0.44\textwidth]{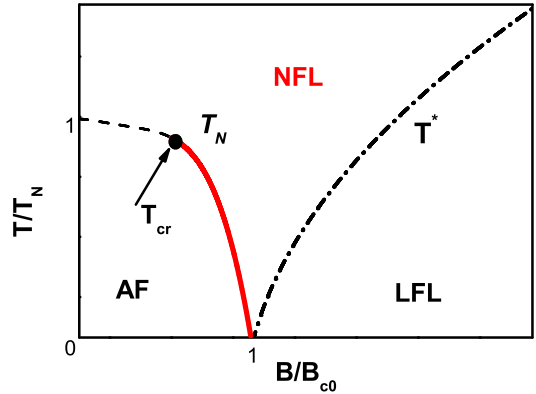}
\vspace*{-0.5cm}
\end{center}
\caption{ Schematic $T-B$ phase diagram for $\rm YbRh_2Si_2$. The
dashed and solid $T_N(B)$ curves separate AF and non-Fermi liquid
(NFL) states representing the field dependence of the N\'eel
temperature. The black dot at $T=T_{cr}$ shown by the arrow  in the
dashed curve is the tricritical temperature, at which the curve of
second order AF phase transitions passes into the curve of the first
ones. At $T<T_{cr}$, the solid line represents the field dependence
of the N\'eel temperature when the AF phase transition is of the
first order. The NFL state is characterized by the entropy $S_{0}$
given by Eq. \eqref{SN}. The dash-dot line separating the NFL state
and the weakly polarized LFL is represented by $T^*(B/B_{c0})\propto
\sqrt{B/B_{c0}}$ \cite{obz}.}\label{fig2}
\end{figure}

In Fig. \ref{fig2}, we present temperature $T/T_N$ versus field
$B/B_{c0}$ schematic phase diagram for $\rm YbRh_2Si_2$. There
$T_N(B)$ is the N\'eel temperature as a function of the magnetic
field $B$. The dash and solid lines indicate boundary of the AF
phase at $B/B_{c0}\leq 1$ \cite{geg}. For $B/B_{c0}\geq 1$, the
dash-dot line marks the upper limit of the observed LFL behavior.
Thus, $\rm YbRh_2Si_2$ demonstrates two different LFL states, where
the temperature-dependent electrical resistivity $\Delta\rho$
follows the LFL behavior $\Delta\rho\propto T^2$, one being weakly
AF ordered ($B\leq B_{c0}$ and $T<T_N(B)$) and the other being
weakly polarized ($B\geq B_{c0}$ and $T<T^*(B)$) \cite{geg}.
At elevated temperatures and fixed magnetic field the NFL regime
occurs which is separated from the AF phase by the curve $T_N(B)$ of
phase transition. In accordance with experimental facts we assume
that at relatively high temperatures $T/T_{N}(B)\simeq 1$ the AF
phase transition is of the second order \cite{TNsteg,geg}. In that
case, the entropy and the other thermodynamic functions are
continuous functions at the curve of the phase transitions $T_N(B)$.
This means that the entropy of the AF phase $S_{AF}(T)$ coincides
with the entropy $S(T)$ of the NFL state
\begin{equation} S_{AF}(T\to T_N(B))=S(T\to
T_N(B)).\label{TN}\end{equation} Since the AF phase demonstrates the
LFL behavior, that is $S_{AF}(T\to 0)\to0$, Eq. \eqref{TN} cannot be
satisfied at diminishing temperatures $T\leq T_{cr}$ due to the
temperature-independent term $S_0$ given by Eq. \eqref{SN}. Thus,
the second order AF phase transition becomes the first order one at
$T=T_{cr}$ as it is shown in Fig. \ref{fig2}. At $T=0$, the critical
field $B_{c0}$ is determined by the condition that the ground state
energy of the AF phase coincides with the ground state energy of the
weakly polarized LFL, and the ground state of $\rm YbRh_2Si_2$
becomes degenerated at $B=B_{c0}$. Therefore, the N\'eel temperature
$T_N(B\to B_{c0})\to 0$.
\begin{figure} [! ht]
\begin{center}
\vspace*{-0.8cm}
\includegraphics [width=0.49\textwidth]{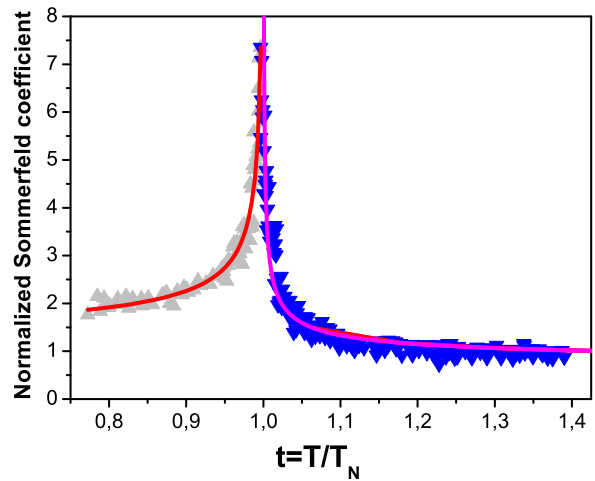}
\end{center}
\vspace*{-0.8cm} \caption{The temperature dependence of the
normalized Sommerfeld coefficient $\gamma_0/A_{+}$ as a function of
the normalized temperature $t=T/T_N(B=0)$ given by the formula
\eqref{TNN} is shown by the solid line. The normalized Sommerfeld
coefficient is extracted from the facts obtained in measurements on
$\rm YbRh_2Si_2$ at the AF phase transition \cite{TNsteg} and shown
by the triangles.}\label{fig3}
\end{figure}

One can expect that the Landau theory of the second order phase
transitions is applicable at the tricritical point $T\simeq T_{cr}$
since the fluctuation theory can lead only to further logarithmic
corrections to the values of the critical indices \cite{land1}. As a
result, upon using the Landau theory we obtain that the Sommerfeld
coefficient $\gamma_0=C/T$ varies as $\gamma_0\propto
1/\sqrt{|t-1|}$ as the tricritical point is approached at fixed
magnetic field \cite{land1}, where $t=T/T_N(B)$. Taking into account
that the specific heat increases in going from the symmetrical to
the asymmetrical AF phase \cite{land1}, we obtain
\begin{equation} \gamma_0(t)=A_{\pm}+\frac{B_{\pm}}{\sqrt{|t-1|}}.\label{TNN}
\end{equation} Here, $B_{\pm}$ are the proportionality factors which
are different for the two sides of the phase transition, the
parameters $A_{\pm}$ related to the corresponding specific heat
$(C/T)_{\pm}$ are also different for the two sides, and ``$+$''
stands for $t>1$, ``$-$'' stands for $t<1$.

The attempt to fit the available experimental data for
$\gamma_0=C(T)/T$ in $\rm YbRh_2Si_2$ at the AF phase transition in
zero magnetic fields \cite{TNsteg} by the function \eqref{TNN} is
reported in Fig. \ref{fig3}. We show there the normalized Sommerfeld
coefficient $\gamma_0/A_{+}$ as a function of the normalized
temperature $T/T_N(B=0)$. It is seen that the normalized Sommerfeld
coefficient $\gamma_0/A_{+}$ extracted from $C/T$ measurements on
$\rm YbRh_2Si_2$ \cite{TNsteg} can be well described by the formula
\eqref{TNN} with $A_{+}=1$.

A few remarks are in order here. The good fitting of the
experimental facts by the function \eqref{TNN} with the critical
exponent $\alpha=1/2$ allows us to predict that the second order AF
phase transition in $\rm YbRh_2Si_2$ changes to the first order
under the application of magnetic field as it is shown by the arrow
in Fig. \ref{fig2}. It is seen from Fig. \ref{fig3} that  at
$t\simeq1$ the peak is sharp, while one would expect that anomalies
in the specific heat associated with the onset of magnetic order are
broad \cite{TNsteg,PTsteg,lohn}. Such a behavior presents
fingerprints that the phase transition is to be changed to the first
order one as it is shown in Fig. \ref{fig2}. As seen form  Fig.
\ref{fig3}, the Sommerfeld coefficient is larger below the phase
transition than above it. This fact is in accord with the Landau
theory stating that the specific heat is increased when passing from
$t>1$ to $t<1$ \cite{land1}.

In summary, we have predicted that the curve of the second order AF
phase transitions in $\rm YbRh_2Si_2$ passes into the curve of the
first order ones at the tricritical point under the application of
magnetic field. This change is generated by the fermion condensation
quantum phase transition. Near the tricritical point the Landau
theory of second order phase transitions is applicable and gives the
critical index $\alpha\simeq1/2$. We demonstrate that this value of
$\alpha$ is in good agreement with the specific heat measurements
and conclude that the critical universality of the fluctuation
theory is violated at the AF phase transition since the second order
phase transition is about to change to the first order one making
$\alpha\to1/2$.

This work was supported in part by the grants: RFBR No. 09-02-00056
and the Hebrew University Intramural Funds. V.R.S. is grateful to
the Lady Davis Foundation for supporting his visit to the Hebrew
University of Jerusalem.

\end{document}